\documentclass[sigconf]{acmart}
\DeclareMathOperator*{\argmax}{arg\,max}
\newcommand{\perm}[2]{{}_{#1}\mathrm{P}_{#2}}

\AtBeginDocument{%
  \providecommand\BibTeX{{%
    \normalfont B\kern-0.5em{\scshape i\kern-0.25em b}\kern-0.8em\TeX}}}

\acmConference[]{Arxiv}
\acmBooktitle{ACM}
\begin{document}

\title{BanditRank: Learning to Rank Using Contextual Bandits}

\author{Phanideep Gampa}
\authornote{Work conducted while the first author was in research internship at Yahoo! JAPAN Research}
\affiliation{\institution{Indian Institute of Technology (BHU) Varanasi}}
\email{gampa.phanideep.mat15@iitbhu.ac.in}

\author{Sumio Fujita}
\affiliation{\institution{Yahoo Japan Corporation}}
\email{sufujita@yahoo-corp.jp}
% \author{Anonymous author}
% \affiliation{\institution{Anonymous institution}}
% \email{aaa@bbb.ccc}

\begin{abstract}
We propose an extensible deep learning method that uses reinforcement learning to train neural networks for offline ranking in information retrieval (IR). We call our method BanditRank as it treats ranking as a contextual bandit problem. In the domain of learning to rank for IR, current deep learning models are trained on objective functions different from the measures they are evaluated on. Since most evaluation measures are discrete quantities, they cannot be leveraged by directly using gradient descent algorithms without an approximation. BanditRank bridges this gap by directly optimizing a task-specific measure, such as mean average precision (MAP), using gradient descent. Specifically, a contextual bandit whose action is to rank input documents is trained using a policy gradient algorithm to directly maximize the reward. The reward can be a single measure, such as MAP, or a combination of several measures. The notion of ranking is also inherent in BanditRank, similar to the current \textit{listwise} approaches. To evaluate the effectiveness of BanditRank, we conducted a series of experiments on datasets related to three different tasks, i.e., web search, community, and factoid question answering. We found that it performs better than state-of-the-art methods when applied on the question answering datasets. On the web search dataset, we found that BanditRank performed better than four strong listwise baselines including LambdaMART, AdaRank, ListNet and Coordinate Ascent.
\end{abstract}




\maketitle

\section{Introduction}
Learning to rank is an important sub-field of information retrieval (IR), which involves designing models that rank documents corresponding to a query in order of their relevance. Considering the type of learning approach used, all ranking models can be classified into three categories, i.e., pointwise, pairwise, and listwise. The ranking models are either trained on indirect objective functions, such as classification related functions, or direct objective functions related to the evaluation measures. Direct optimization of IR measures has been a long standing challenge in the learning-to-rank domain. If we only consider bounded IR measures such as MAP, a theoretical justification is provided regarding the superiority of direct optimization techniques~\citep{qin2010general}. That study states that if an algorithm  can directly optimize an IR measure on the training data, the ranking function learned with the algorithm will be one of the best ranking functions one can obtain in terms of expected test performance with respect to the same IR measure. Several algorithms have been developed that use direct optimization, and they can be grouped into three categories. The algorithms in the first category try to optimize the surrogate objective functions, which are either upper bounds of IR measures~\citep{xu2007adarank,yue2007support,chapelle2007large} or smooth approximations of IR measures~\citep{guiver2008learning,taylor2008softrank}. The algorithms in the second category smoothly approximate the true gradient of the evaluation measures, similar to LambdaRank~\citep{burges2007learning,donmez2009local,burges2010ranknet,yuan2016lambdafm}. The algorithms in the third category directly optimize evaluation measures in the form of rewards without any approximation using reinforcement learning such as MDPRank~\citep{wei2017reinforcement,zhao2018reinforcement}. However, except for some algorithms like LambdaRank, most of the algorithms in all the categories are only suitable for models with less parameters~\citep{donmez2009local} making it difficult to use deep neural networks, which are quite effective.

Deep learning~\citep{lecun2015deep} models have been proven to be effective with state-of-the-art results in many machine learning applications such as speech recognition, computer vision, and natural language processing, which leads to the introduction of neural networks in IR. Neural networks have been used for functions such as automatic feature extraction and comparison and aggregation of local relevance~\citep{guo2016deep,hu2014convolutional,huang2013learning,wan2016match,pang2017deeprank}. But, the neural networks are generally trained on objective functions such as cross entropy, which is not related to the evaluation measures. They do not have information about the measures that they are going to be evaluated on, i.e., the objective functions indirectly optimize the  evaluation measures. Since most evaluation measures such as MAP, mean reciprocal rank (MRR), and normalized discounted cumulative gain (nDCG) are not differentiable, they cannot be used as the objective functions for training the neural networks.  

For leveraging the efficacy of neural networks and superiority of direct optimization, we propose an extensible deep learning method called BanditRank. BanditRank formulates ranking as a contextual bandit problem and trains neural networks using the policy gradient algorithm~\citep{sutton2000policy}, for directly maximizing the target measures. Contextual bandit is a type of reinforcement learning algorithm used in decision-making scenarios in which an action has to be taken by an agent depending on the provided context. The exact details of the formulation are provided in Section ~\ref{sec:formulation}. BanditRank follows the listwise approach by treating a query and the corresponding candidate documents as a single instance for training. BanditRank is extensible in the sense that it provides a methodology for training neural networks using reinforcement learning for ranking. Therefore, it can be used with any text-matching architecture for feature extraction and jointly trained or it can use the features extracted from a pre-trained model. For example, the LETOR 4.0 dataset~\citep{qin2013introducing} provides 46-dimensional feature vectors corresponding to each query-document pair that can be leveraged directly for training. Since BanditRank is a deep learning method, it is also extensible with respect to the choice of architectures that can be used. We focused on offline ranking tasks in which external relevance labels are provided for training.

Empirically, we prove that BanditRank is superior when compared to other strong baselines. We conducted a series of experiments on three datasets in the domains of question answering and web search. The major contributions of this paper are summarized as follows:  
\begin{itemize}
\item For training neural networks by directly optimizing evaluation measures using gradient descent algorithms, we formulate the ranking problem as a contextual bandit and introduce a new deep learning method called \textit{BanditRank} for ranking. 
\item To the best of our knowledge, BanditRank is the first listwise deep learning method that uses reinforcement learning to train neural networks for offline ranking purposes. We enabled this by introducing a hybrid training objective in order to solve the exploration problem when the number of possible actions is large. 
\item BanditRank provided state-of-the-art results when applied on both InsuranceQA~\citep{feng2015applying} and WikiQA~\citep{yang2015wikiqa} datasets outperforming the previous best method at the time of writing of this paper. 
\item In the web-search task, when applied on the benchmark MQ2007 ~\citep{qin2013introducing} dataset using only the provided 46-dimensional features, BanditRank achieved better results than the state-of-the-art learning-to-rank algorithm LambdaMART~\citep{burges2010ranknet} and clearly outperformed other listwise baselines such as CoordinateAscent~\citep{metzler2007linear}, ListNet~\citep{cao2007learning}, and AdaRank~\citep{xu2007adarank}.
\end{itemize}

The remainder of the paper is structured as follows. In the next section, we briefly discuss related studies. In Section~\ref{sec:formulation}, we give the formulation of ranking as a contextual bandit. In Section~\ref{sec:model architecture}, we provide the details of the model architecture used for our experiments. We explain the experiments we conducted along with a comparative study of rewards in Section~\ref{sec:experiments}. We conclude the paper in Section~\ref{sec:conclusion}.

\section{Related Work}
BanditRank is similar to BanditSum \cite{dong2018banditsum}, which was proposed earlier for extractive summarization tasks in NLP. BanditSum introduces a theoretically grounded method based on contextual bandit formalism for training neural-network-based summarizers with reinforcement learning. We have adapted the formulation of ranking as a contextual bandit from that of BanditSum. Adaptation of the contextual bandit framework to the ranking problem is not straightforward at all, for example, a naive application of BanditSum suffers from inadequate exploration when the number of actions is very large which is prevalent in ranking tasks. Thus we propose the use of hybrid loss for leveraging the feedback from a supervised loss function as explained in Section~\ref{subsec: hybrid training objective}. Reinforcement learning was used for directly optimizing measures such as BLEU~\citep{papineni2002bleu} and ROUGE~\citep{lin2004rouge} in different tasks of natural language processing such as summarization and sequence prediction~\citep{paulus2017deep,lee2017automatic,ranzato2015sequence,bahdanau2016actor}. 

In the domain of learning-to-rank for IR, MDPRank~\citep{wei2017reinforcement} uses reinforcement learning for ranking by formulating ranking as a sequential decision process. Since the sequential models are affected by the order of the decisions, they may be biased towards selecting documents with low relevance level at the beginning. MDPRank is not suitable for training neural networks because a model with only 46 weight parameters requires more than 10000 epochs for convergence. In contrast, BanditRank is suitable for deep architectures, and all the best results of BanditRank were achieved in less than 30 epochs of training. Another issue with the setting of MDPRank is that the number of possible rankings for a query $q$ with $n_q$ number of candidate documents is $n_q!$, which is quite large making exploration more difficult. In contrast, BanditRank has more flexibility and freedom to explore the search space, as it samples a fixed number of documents $M$ without replacement based on the affinity scores during training, reducing the search space to $\perm{n_q}{M}<< n_q!$ for small $M$. This is because BanditRank uses listwise approach by treating all the candidate documents corresponding to a query as a single state. The policy gradient algorithm was also used to train the generator of IRGAN~\citep{wang2017irgan}, but the rewards for the generator depend on the scoring function learned by the discriminator. The training of IRGAN is similar to that of SeqGAN~\citep{yu2017seqgan}, which is based on the idea of using the policy gradient algorithm for tackling the generator differentiation problem due to the discrete outputs produced by the generator. Both Bandits~\citep{radlinski2008learning,kveton2015cascading,katariya2016dcm} and MDPs~\citep{zeng2018multi} were used to model the interactive process between a search engine and user with the user providing implicit relevance feedback. An overview of approaches that use reinforcement learning for different IR tasks such as query reformulation, recommendation and session search can be found in a previous paper~\citep{zhao2018reinforcement}. BanditRank's action is similar to the formulation of ListNet~\citep{cao2007learning}, which is based on the permutation of the input documents. However, both approaches differ with respect to the training method and structure of the probability model used.

\section{BanditRank Formulation }
\label{sec:formulation}
We formulate ranking as a contextual bandit trained using policy gradient reinforcement learning. A bandit is a decision-making algorithm in which an agent repeatedly chooses one out of several actions and receives a reward based on this choice. The goal of the agent is to maximize the cumulative reward it achieves by learning the actions that yield good rewards. The term \textit{agent} is generally used to refer to an entity or model that interacts with the environment. Contextual Bandit is a variant of the bandit problem that conditions its action on the context or state of the environment and observes the reward for the chosen action only. It forms a subclass of Markov decision processes with the length of each episode being one. Formally, assume there is an environment with context space $X$ and action space $A$. The agent interacts with the environment in a series of time steps. At each time step $t$, the agent observes a context $x_{t}\in X$, chooses an action $a_t \in A$, and observes a reward for that action $r(a_{t})$. The goal of the agent is to maximize the cumulative rewards it achieves over a certain period.

Now, we can formulate the ranking problem as a contextual bandit with the environment being the dataset of queries and documents. The set of query-document pairs corresponding to a single query is treated as a context, and each permutation of the candidate documents is treated as a different action. Formally, given a query $q$ and its candidate documents $d=\{d_1,d_2,\ldots,d_{n_q}\}$, each context is the set $c$ given by $c=\{(q,d_1),(q,d_2),\ldots,(q,d_{n_q})\}$. Where $n_q$ is the number of candidate documents of $q$, and the cardinality of $c$ is given by $n_c=n_q$. Given $c$, the action of the agent is given by the permutation $a_c=(d_{k_1},d_{k_2},\ldots,d_{k_{n_q}})$ of the candidate documents, where $k_t\in\{1,2,\ldots,n_q\}$ and $k_t \neq k_{t'}$ for $t\neq t'$. The reward is given by a scalar function $R(a_c,g_c)$ that takes action $a_c$ and the ground-truth permutation $g_c$ corresponding to $c$ as the input. The $g_c$ is nothing but the candidate documents sorted in descending order according to their relevance levels. The notation $R$ is a scalar reward function defined using a combination of measures such as MAP and MRR.

The action taken by the agent is determined by its \textit{policy}. In the current formulation, a policy is a neural network $p_{\theta}(.|c)$ parameterized by $\theta$. For each input $c$, $p_{\theta}(.|c)$ encodes a probability distribution over permutations of the candidate documents. The goal is to find $\theta$ that cause the network to assign high probability to the permutations, which can yield good rewards induced by $R$. This can be achieved by maximizing the following objective function with respect to $\theta$ :
\begin{equation}
\label{eq:objective function}
J(\theta)=E[R(a_c,g_c)],
\end{equation}
where the expectation is taken over $c$ paired with $g_c$ and $a_c$ generated according to $p_{\theta}(.|c)$. The above objective function is a standard objective function used in the reinforcement-learning domain, which maximizes the expected reward. The negative of the expectation can be treated as the loss function.

\subsection{Structure of Policy $p_{\theta}(.|c)$}
\label{subsec:structure of p}
The exact action of the agent depends on the chosen structure of $p_{\theta}(.|c)$. We follow the approach used for extractive summarization~\citep{dong2018banditsum} because of its simplicity and effectiveness. With this approach, $p_{\theta}(.|c)$ is decomposed into a deterministic function $\pi_{\theta}$, which contains all the network's parameters, and $\mu$, a probability distribution induced by the output of $\pi_{\theta}$ defined as
\begin{equation}
\label{eq: structure of p}
p_\theta(.|c)=\mu(.|\pi_\theta(c))
\end{equation}   

Provided a $c$ corresponding to a $q$, the network $\pi_\theta$ outputs a real valued vector of document affinities within the range $[0,1]$. The length of the vector is equal to the number of candidate documents $n_c$ corresponding to $q$, i.e., $\pi_\theta(c) \in \mathbb{R}^{n_c} $. The affinity score of a document $d_i$ given by $\pi_\theta(c)_i$ represents the network's propensity to keep the document at the top position in the output permutation. Specifically, the interpretation of the affinity scores is highly dependent upon the type of reward signal used. For example, if only Precision@1 is used as the reward signal, the focus of the network would mainly be on the permutations that contain a relevant document at the first position. 

Provided the above document affinities $\pi_\theta(c)$, $\mu$ implements a process of repeated sampling without replacement by repeatedly normalizing the set of affinities of documents not yet selected. In total, $M$ unique documents are sampled yielding an ordered subset of the candidate documents. For exploring the action space, a small probability $\epsilon$ of sampling uniformly from all remaining documents is included at each step of the sampling. This is similar to the $\epsilon$-greedy technique generally used in the reinforcement learning problems for exploration. According to the prescribed definition of $\mu$, the probability $p_\theta(a_c|c)$ of producing a permutation $a_c$ corresponding to $c$ according to \eqref{eq: structure of p} is given by
\begin{equation}
\label{eq: expression of p}
%p_\theta(a|c)=\prod_{i=1}^{M}\left( \frac{\epsilon}{n_c-i+1}+\frac{(1-\epsilon)\pi_\theta(c)_i}{z(c)-\sum_{k=1}^{i-1}\pi_\theta(c)_k} \right),
p_\theta(a_c|c)=\prod_{i=1}^{M}\left( \frac{\epsilon}{n_c-i+1}+\frac{(1-\epsilon)\pi_\theta(c)_{k_i}}{z(c)-\sum_{l=1}^{i-1}\pi_\theta(c)_{k_l}} \right),
\end{equation} 
where $k_t$ is the index to the $t$-th document in $a_c$, $d_{k_t}$ and $z(c)=\sum_{m=1}^{n_c} \pi_\theta(c)_m$. 
We define $M=min(n_c,M')$, where $M'$ is an integer hyper parameter that depends on the environment or dataset. This sampling method with exploration is followed only during training time. At test time, we output all the candidate documents sorted in descending order according to their affinity scores.

\subsection{Policy Gradient Reinforcement Learning}
\label{subsec:policy gradient}
The gradient of the objective function \eqref{eq:objective function} cannot be calculated directly as $a_c$ is discretely sampled while calculating $R(a_c,g_c)$. This is a common situation in most reinforcement-learning tasks. However, the gradient of the objective function can be calculated after a reformulation of the expectation term according to the REINFORCE algorithm~\citep{williams1992simple,sutton2000policy}. It tells us that the gradient of that function can be calculated using the following equation:
\begin{equation}
\label{eq:reinforce gradient}
\nabla_\theta J(\theta)=E[\nabla_\theta \log{p_\theta(a_c|c)}R(a_c,g_c)],
\end{equation}   
where the expectation is over the same variables as \eqref{eq:objective function}.

Given a context-true permutation pair $(c,g_c)$ sampled from the dataset or environment $D(c,g_c)$, the gradient can be derived using the following reformulation of the expectation in \eqref{eq:objective function}:
\begin{align}
\nabla_\theta J(\theta) &= \nabla_\theta E[R(a_c,g_c)]  \nonumber  \\
&= \nabla_\theta E_{(c,g_c)\sim D(c,g_c),a_c\sim p_\theta(.|c)}[R(a_c,g_c)]  \nonumber  \\
&=\sum_{(c,g_c)\sim D(c,g_c)} \sum_{i}\nabla_\theta p_\theta(a_c^i|c)R(a_c^i,g_c) \label{derivation: step 2}   \\
&= \sum_{(c,g_c)\sim D(c,g_c)}\sum_{i}p_\theta(a_c^i|c)\nabla_\theta \log{p_\theta(a_c^i|c)}R(a_c^i,g_c)\label{derivation: step 3}  \\
&=  E_{(c,g_c)\sim D(c,g_c),a_c\sim p_\theta(.|c)}[\nabla_\theta \log{p_\theta(a_c|c)}R(a_c,g_c)] \nonumber  \\
&= E[\nabla_\theta \log{p_\theta(a_c|c)}R(a_c,g_c)] \nonumber
\end{align}
Step \eqref{derivation: step 2} follows from the definition of expectation for discrete quantities and the linearity of the gradient operator. Step \eqref{derivation: step 3} is the reformulation of the gradient term, which is an important step in the derivation initially given by the REINFORCE~\citep{williams1992simple} algorithm. The expectation in \eqref{eq:reinforce gradient} is empirically calculated by first sampling a context-true permutation pair $(c,g_c)$, sampling $B$ permutations $a_c^1,a_c^2,\ldots,a_c^B$ from $p_{\theta}(.|c)$ using the sampling method mentioned in Section ~\ref{subsec:structure of p}, and finally taking the average. Empirically, the inner expectation of \eqref{eq:reinforce gradient} is given by
\begin{equation}
\label{eq:gradient approx}
\nabla_\theta J_c(\theta)\approx \frac{\sum_{i=1}^B \nabla_\theta \log{p_\theta(a_c^i|c)}R(a_c^i,g_c)}{B}
\end{equation}
The number $B$ is also an integer hyperparameter that mainly depends on the dataset. Given the expression for $p_\theta(a_c|c)$ \eqref{eq: expression of p}, the gradient \eqref{eq:gradient approx} can be calculated by any automatic differentiation package. As mentioned in Section~\ref{subsec:structure of p}, we sample $M=min(n_c,M')$ number of documents from the candidate documents during training time. Therefore, we take reward feedback from an $M$-length ordered subset. Since we cannot efficiently explore the whole action space for large $M$ as the number of possible actions or permutations would then become $\perm{n_c}{M}$ ~\footnote{Permutation $\perm{n}{r}$ is an increasing function of $r$.}, we choose $M$ based on the average number of relevant documents per query in the dataset. The $B$ determines the exact number of actions we explore for each context during each epoch, which can be seen from the approximation in \eqref{eq:gradient approx}. In our experiments, we obtained very good results even-though $B$ was set to be small, i.e., $B=20$ or $B=30$.

The gradient estimate in \eqref{eq:gradient approx} is prone to have high variance~\citep{sutton2000policy}. Moreover, all target measures, such as MAP, MRR, and nDCG, are always non negative, which increases the probability of every sampled permutation according to the objective function. We would prefer the probability of a bad permutation in terms of reward should be decreased. We use a baseline function, which is subtracted from all rewards. This decreases the variance of the estimate by acting as an advantage function, and it ensures that the permutations with low rewards receive negative rewards. If chosen appropriately, the advantage function can significantly reduce the variance of the estimate~\citep{sutton2000policy} without biasing the estimate. Using a baseline $r_{base}$, the sample-based estimate \eqref{eq:gradient approx} becomes
\begin{equation}
\label{eq:baseline gradient}
\nabla_\theta J_c(\theta)\approx \frac{\sum_{i=1}^B \nabla_\theta \log{p_\theta(a_c^i|c)}[R(a_c^i,g_c)-r_{base}]}{B}
\end{equation}    

For choosing the baseline function, we follow the terminology of self-critical reinforcement learning, in which the test time performance of the current model is used as the baseline~\citep{dong2018banditsum,ranzato2015sequence,paulus2017deep,rennie2017self}. Therefore, while calculating the gradient estimate \eqref{eq:baseline gradient} after sampling the context-true permutation pair $(c,g_c)$, we greedily generate a permutation using the current model similar to the test time action.
\begin{equation}
a_c^{greedy} = \argmax_{a_c} p_\theta(a_c|c)
\end{equation}
The baseline for a $c$ is then calculated by setting $r_{base}=R(a_c^{greedy},g_c)$. Therefore, all the permutations with reward greater than the \textit{greedy} permutation receive positive rewards and other permutations receive negative rewards. The baseline is also intuitive in the way that it is different for different contexts.

\subsection{Reward Function $R$}
As mentioned earlier, the reward function can be a single target measure or a combination of several measures. For the question answering datasets, the following reward function was used:
\begin{equation}
\label{eq:reward 1}
R(a_c,g_c) = \frac{AP(a_c,g_c)+RR(a_c,g_c)}{2}
\end{equation}
For the web search dataset, the following reward function was used:
\begin{equation}
\label{eq:reward 2}
R'(a_c,g_c) = \frac{AP(a_c,g_c)+nDCG@10(a_c,g_c)}{2},
\end{equation}
where the measures average precision (AP), reciprocal rank (RR), and nDCG@10 are traditional IR measures. In the experiments section, we also provide a simple comparison of different reward functions on the web search dataset. 

\subsection{Hybrid Training Objective}
\label{subsec: hybrid training objective}
As mentioned in the Section~\ref{subsec:policy gradient}, the problem of exploring the action space when $M$ is large can be tackled using a hybrid loss, which is a combination of the reinforcement learning loss and a standard supervised learning loss such as binary cross entropy. The supervised loss can guide the training initially when the exploration by the model is in the starting stages for large $M$. Even though the number of actions explored with the model at each epoch given by $B$ is small for large $M$, i.e., $B << \perm{n_c}{M}$, a supervised signal can help the model by compensating the loss incurred due to the inefficient exploration. The hybrid loss function is given as follows:
\begin{equation}
\label{eq:hybdrid loss}
L_{hybrid} = \gamma L_{rl}+ (1-\gamma)L_{sl}
\end{equation}
where $L_{rl}$ is the loss given by the reinforcement-learning algorithm, which is the negative of \eqref{eq:objective function}, and $L_{sl}$ is a supervised loss such as binary cross entropy. The notation $\gamma$ is a scaling factor accounting for the difference in magnitude between $L_{rl}$ and $L_{sl}$. It is a hyperparameter lying between 0 and 1. We found the hybrid loss to be effective in the case of the web search dataset where the average number of relevant documents per query was equal to 10.3. Since we use binary cross entropy as the supervised loss, the hybrid training objective is a blend of the \textit{pointwise} objective function $L_{sl}$ and a \textit{listwise} objective function $L_{rl}$. The hybrid training objective still has direct control over the target measures weighted by $\gamma$. Similar hybrid loss was used in the domain of NLP in some papers, e.g.,~\citep{paulus2017deep,wu2016google}.

\section{Model Architecture}
\label{sec:model architecture}
In this section, we discuss the neural architecture $\pi_\theta$ we used in our experiments. For demonstrating the extensibility of BanditRank, we considered two scenarios for the experiments and show that BanditRank performs well in both the scenarios. 

\subsection*{Scenario 1}
\label{subsec: scenario1}
In this scenario, we decomposed $\pi_\theta$ into two neural network architectures $f_{\theta1}$ and $b_{\theta2}$; $f_{\theta1}$ for extracting the feature vectors from raw texts of query or documents and $b_{\theta2}$ for a bandit network that yields the document affinities. For $f_{\theta1}$, since any text-matching neural network~\citep{tay2018multi,wang2016compare,bian2017compare} that can provide a single feature vector corresponding to each query-document pair was suitable, we have used the architecture similar to the recently proposed Multi Cast Attention Networks (MCAN)~\citep{tay2018multi}. The $b_{\theta2}$ architecture was chosen to be a simple feed forward neural network with an output sigmoid unit for yielding the document affinities corresponding to a query. While training, the two architectures were treated as a single architecture by positioning the bandit network on top of the text-matching network. Formally, provided with a context $c=\{(q,d_1),(q,d_2),\ldots,(q,d_{n_q})\}$, we passed it through both networks to obtain the document affinities $\pi_\theta(d)$ as given below:
\begin{align}
&f_{\theta1}(c)=c_1,c_2,\ldots,c_{n_q} \nonumber \\
&b_{\theta2}(c_1,c_2,\ldots,c_{n_q}) = \pi_\theta(d) \nonumber
\end{align}
where $c_i$ is a feature vector corresponding to the query-document pair $(q,d_i)$.

\subsection*{Scenario 2}
\label{subsec: scenario2}
In this scenario, we have only used $b_\theta$ for training. The feature vectors corresponding to the query and document texts were extracted using a pre-trained neural language model such as BERT~\citep{devlin2018bert}, which is a state-of-the-art unsupervised language model in NLP. In web search datasets, such as MQ2007~\citep{qin2013introducing}, 46-dimensional feature vectors composed of basic features including BM25, term frequency (TF), and LMIR, are provided for each query-document pair. We directly use those vectors while conducting experiment on MQ2007 dataset. Formally, provided with the query-document feature vectors $c_1,\ldots,c_{n_q}$ corresponding to a context $c=\{(q,d_1),\ldots,(q,d_{n_q})\}$, we have obtained the document affinities $\pi_\theta(d)$ as given below :
\begin{align*}
b_{\theta}(c_1,c_2,\ldots,c_{n_q}) = \pi_\theta(d)
\end{align*}
These scenarios are intended for separating the functionality of the text-matching and bandit networks. Therefore, the inputs and outputs are not necessarily in the above prescribed format.

The results obtained in both the scenarios indicate that the bandit network is not entirely dependent on a text-matching network for providing good results. The exact details of the scenarios and the neural network architectures used for each dataset are provided in the experiments section~\footnote{We will make our implementation publicly available when the paper is accepted.}.

\section{Experiments}
\label{sec:experiments}
We conducted our experiments on three different datasets in the domains of question answering and web search. For the question answer ring task, we tested BanditRank on InsuranceQA ~\citep{feng2015applying}, which is a community question answering dataset (closed domain), and on WikiQA~\citep{yang2015wikiqa}, which is a well studied factoid question answering dataset (open domain). For the web search task, we conducted our experiments on the benchmark MQ2007~\citep{qin2013introducing} dataset. Since the baselines and evaluation measures are different for the above three datasets, we divide this section into three subsections each dealing with a specific dataset. For each dataset, we provide the details of the architecture used, implementation details, details of the baselines, and obtained results. We adopted Scenario 1 for the InsuranceQA dataset and Scenario 2 for the MQ2007 dataset and compared both scenarios on the WikiQA dataset. Finally, we conducted a comparative study on different reward choices for training on the MQ2007 dataset. We choose the exploration probability mentioned in Section~\ref{subsec:structure of p} as $\epsilon=0.1$ for all the experiments. 

\subsection{Experiment 1: InsuranceQA}
\subsubsection{Dataset and Evaluation Measures}
InsuranceQA~\citep{feng2015applying} is a well studied community question answering dataset ~\footnote{https://github.com/shuzi/insuranceQA} with questions submitted by real users and answers composed by professionals with good domain knowledge. In this task, BanditRank was expected to select the correct answer among a pool of candidate answers with negative answers randomly sampled from the whole dataset. The dataset consists of two test datasets for evaluation. The evaluation measure for this task was Precision@1. The statistics of the dataset along with the average number of relevant answers per question are given in Table ~\ref{table: insuranceqa}.
\begin{table}[h]
\caption{Statistics of InsuranceQA dataset}
\label{table: insuranceqa}
\begin{tabular}{l c c c}
\hline
Split & \# of questions & \# of correct answers & \# of avg-rel-per-q \\
\hline
train & 12887 & 18540 & 1.43 \\
dev & 1000 & 1454 & 1.45 \\
test1 & 1800 & 2616 & 1.45 \\
test2 & 1800 & 2593 & 1.44 \\
\hline
\end{tabular}
\end{table}

\subsubsection{Model and Implementation Details} 
We used two architectures for Scenario 1. For the text matching part, we used the architecture of the recently proposed Multi Cast Attention Network (MCAN)~\citep{tay2018multi}. Specifically, we used the same architecture until the mean max pooling operation layer of the MCAN, which provides a fixed dimensional feature vector of each question or answer sentence. We used the sum function (SM) as the compression function for casting the attention. For the bandit network, we modified the prediction layer of MCAN with a sigmoid unit in the output layer and kept the two highway layers~\citep{srivastava2015highway} intact. We conducted the experiments by replacing the highway networks with simple neural networks with the same dimensions, but the best results were obtained when highway layers were used. Highway networks~\citep{srivastava2015highway} are gated nonlinear transform layers that control the information flow similar to the LSTM layers~\citep{hochreiter1997long} for sequential tasks.

We randomly initialized the embedding matrix with 100 dimensional vectors sampled from standard normal distribution and fixed them during training. The hidden size of the LSTM layers was set to 300. The dimensions of the two highway prediction layers were set to 200 with ReLU being the activation function. Once we obtained the feature vectors corresponding to the question and answer sentences from the text-matching network as $h_q,h_a\in \mathbb{R}^{600}$, we passed the following vector $h_{qa}=[h_q;h_a;h_q\odot h_a;h_q-h_a] \in \mathbb{R}^{2400} $ to the bandit network, where $\odot$ is the pointwise multiplication operation and $[.;.]$ is a vector concatenation operator. The vectors $h_{qa}$ correspond to the query-document feature vectors $c_i$ mentioned in Section~\ref{subsec: scenario1}. Before passing $h_{qa}$ to the highway layers of bandit network, we used a single feed forward layer with a ReLU activation function for projecting $h_{qa}$ into a 200-dimensional space. A dropout of 0.2 was applied to all layers except the embedding layer. The sequences were padded to their batch-wise maximums. We optimized the model using the Adam optimizer~\citep{kingma2014adam} with the beta parameters set to $(0.9,0.999)$ and a weight decay of $1e^{-6}$ was used for regularization. We used the hybrid training objective defined in Eq.~\eqref{eq:hybdrid loss} with $\gamma$ tuned over the set of values $[0.5,0.75,1]$. An initial learning rate of $5e^{-5}$ was used for BanditRank with $\gamma=1$ and $1e^{-4}$ for BanditRank with other $\gamma$ values. We set $M',B$ to $M'=5$ and $B=20$ for calculating the gradient in Eq.~\eqref{eq:baseline gradient}. For BanditRank with $\gamma=0.75$, $M'$ was set to 20. We used the reward function defined as Eq.~\eqref{eq:reward 1} during training. 
\begin{table}[h]
\caption{Precision@1 for InsuranceQA dataset. Best results are in bold and second best are underlined.}
\label{table: insurance results}
\begin{tabular}{l c c}
\hline
& test-1 & test-2 \\
\hline
IR model~\citep{bendersky2010learning} & 0.551 & 0.508 \\
QA-CNN~\citep{santos2016attentive} & 0.6133 &0.5689 \\
LambdaCNN~\citep{santos2016attentive,zhang2013optimizing} & 0.6294 &0.6006 \\
IRGAN~\citep{wang2017irgan} & 0.6444 & 0.6111 \\
CNN with GESD~\citep{feng2015applying} & 0.653 & 0.61 \\
Attentive LSTM~\citep{tan2016improved} & 0.69 & 0.648 \\
IARNN-Occam ~\citep{wang2016inner} & 0.689 & 0.651 \\
IARNN-Gate~\citep{wang2016inner} & 0.701 & 0.628 \\
Comp-Agg(MULT)~\citep{wang2016compare} & 0.752 & 0.734 \\
Comp-Agg(SUBMULT+NN)~\citep{wang2016compare} & 0.756 & 0.723 \\
\hline
BanditRank($\gamma=1$) & \underline{0.8494} & \underline{0.8283} \\
BanditRank($\gamma=0.75$) & \textbf{0.8572} & \textbf{0.8522} \\
\hline
\end{tabular}
\end{table}

\subsubsection{Baselines and Results} 
We compared the performance of BanditRank against all current methods that achieved significant results on this dataset. The competitive baselines are the IR model~\citep{bendersky2010learning}, CNN with GESD (from the authors who created the InsuranceQA dataset)~\citep{feng2015applying}, Attentive LSTM~\citep{tan2016improved}, IARNN-Occam ~\citep{wang2016inner}, IARNN-Gate~\citep{wang2016inner}, QA-CNN~\citep{santos2016attentive}, LambdaCNN~\citep{santos2016attentive,zhang2013optimizing}, IRGAN~\citep{wang2017irgan}, and the method with the previous best P@1 measure, Comp-Agg~\citep{wang2016compare}. A description of all the baselines can be found in previous studies~\citep{wang2016compare,wang2017irgan}. As the testing splits were the same for all methods, we report the P@1 measures directly from those studies.

The results in Table ~\ref{table: insurance results} indicate the superiority of BanditRank over all other methods. BanditRank with $\gamma=1$ achieved the second best P@1 measure, this is equivalent to training the model only with the reinforcement loss. The results further improved using a hybrid loss with $\gamma=0.75$ weight to the reinforcement loss $L_{rl}$. Therefore, providing some weight to the supervised loss improved the performance of BanditRank. BanditRank exhibited significant improvement in P@1 measure by 13.3\% on the test-1 dataset and 16.1\% on the test-2 dataset when compared to the previous best method. 

\subsection{Experiment 2: WikiQA}
\subsubsection{Dataset and Evaluation Measures} 
WikiQA~\citep{yang2015wikiqa} is a well-known open domain question answering dataset in contrast to InsuranceQA, which is a closed domain question answering dataset. The dataset was constructed from real queries on Bing and Wikipedia. In this task, the models were expected to rank the candidate answers according to the question. The evaluation measures for this task were MAP and MRR. The statistics of the dataset are given in Table ~\ref{table: wikiqa}. 
\begin{table}[h]
\caption{Statistics of WikiQA dataset}
\label{table: wikiqa}
\begin{tabular}{l c c c}
\hline
Split & \# of questions & \# of correct answers & \# of avg-rel-per-q \\
\hline
train & 873 & 1040 & 1.19 \\
dev & 126 & 140 & 1.11 \\
test & 243 & 293 & 1.20 \\
\hline
\end{tabular}
\end{table}

\subsubsection{Model and Implementation details}
We considered both the scenarios given in Section~\ref{sec:model architecture} for WikiQA dataset. For Scenario 1, we used the same setting of MCAN as InsuranceQA. The only difference was the type of embedding used. We initialized the embedding matrix with 300-dimensional GloVe embeddings~\citep{pennington2014glove} and fixed them during training. We used the reward function defined as Eq.~\eqref{eq:reward 1} during training. A dropout of 0.2 was applied to all layers except the embedding layer. The sequences were padded to their batch-wise maximums. We optimized the model using the Adam optimizer~\citep{kingma2014adam} with the beta parameters set to $(0.9,0.999)$, and a weight decay of $1e^{-6}$ was used for regularization. We used the hybrid training objective defined in Eq.~\eqref{eq:hybdrid loss} with $\gamma$ tuned over the set of values $[0.25,0.5,0.75,1]$. We provide the results of the two best performing models with respect to $\gamma$. An initial learning rate of $5e^{-5}$ was used for BanditRank with $\gamma=1$ and $1e^{-4}$ for BanditRank with other $\gamma$ values. We set $M',B$ to $M'=3$ and $B=20$ for calculating the gradient in Eq.~\eqref{eq:baseline gradient}. The $M'$ was chosen according to the average number of relevant queries, which is much less for the WikiQA dataset. 

For Scenario 2, we extracted word-level ~\footnote{Please note that, even-though only word level features are extracted from BERT, these features encode contextual information of both the input sentences similar to that of a text matching network.} features from a pre-trained~\footnote{https://github.com/huggingface/pytorch-pretrained-BERT} BERT~\citep{devlin2018bert} language model, which takes a question-answer sentence pair $(q,a)$ as the input. There are two versions of the BERT model available, BERT-base and BERT-large. We conducted our experiments on features extracted with both versions. We used the concatenation of word-level features obtained from the last four layers of the BERT language model for training. BERT-base produced 3072-dimensional feature vectors for each word in the input sentence while BERT-large produced 4096-dimensional feature vectors after the concatenation. These contextual word embeddings were passed through a two-layer bidirectional LSTM layer followed by mean pooling for obtaining a fixed dimensional representation ~\footnote{Although LSTM layers are used before the bandit network, we include this in Scenario 2 as the true text matching was carried out with the BERT model, LSTM layers were only used to obtain a fixed dimensional representation of the question-answer pair.} of $(q,a)$. These vectors correspond to the $c_i$ vectors mentioned in Section~\ref{subsec: scenario2}. Regarding the architecture of the bandit network, we chose a feed forward network with a single hidden layer followed by a sigmoid unit at the output layer. Tanh was used as the activation function. For the BanditRank method trained on features extracted from BERT-base, we set the dimensions of the LSTM layer to 768. For BERT-large, we set the dimensions of the LSTM layer to 1024. For the feed forward layer, we set the dimensions of the hidden layer to 256 for both type of features. A dropout of 0.4 was applied to all layers. We optimized BanditRank using the Adam optimizer~\citep{kingma2014adam} with the beta parameters set to $(0,0.999)$ and used a weight decay of $1e^{-6}$ for regularization. We used the hybrid training objective defined in Eq.~\eqref{eq:hybdrid loss} with $\gamma$ tuned over the set of values $[0.5,0.75,1]$. We provide the results of the best performing model with respect to $\gamma$. An initial learning rate of $8e^{-5}$ was used for BanditRank with $\gamma=1$ and $1e^{-4}$ for the BanditRank method with other $\gamma$ values. We set $M',B$ to $M'=5$ and $B=20$ for calculating the gradient in Eq.~\eqref{eq:baseline gradient}. We used the reward function defined with Eq.~\eqref{eq:reward 1} during training. 
\begin{table}[h]
\caption{Test-set performance on WikiQA dataset. Best results are in bold and second best are underlined.}
\label{table: wikiqa results}
\begin{tabular}{l c c}
\hline
& MAP& MRR \\
\hline
CNN-Cnt~\citep{yang2015wikiqa} & 0.652 & 0.665 \\
QA-CNN~\citep{santos2016attentive} & 0.689 &0.696 \\
NASM~\citep{miao2016neural} & 0.689 &0.707 \\
Wang et al~\citep{wang2016sentence} & 0.706 & 0.723 \\
He and Lin~\citep{he2016pairwise} & 0.709 &0.723 \\
NCE-CNN~\citep{rao2016noise} & 0.701 & 0.718\\
BIMPM~\citep{wang2017bilateral} & 0.718 & 0.731 \\
Comp-Agg~\citep{wang2016compare} & 0.743 & 0.755 \\
Comp-Clip~\citep{bian2017compare} & \underline{0.754} & \underline{0.764} \\
\hline
Scenario 1 & & \\
BanditRank($\gamma=0.75$) & 0.6663 & 0.673 \\
BanditRank($\gamma=1$) & 0.7043 & 0.716 \\
\hline
Scenario 2 & & \\
BanditRank-BERT-base ($\gamma =1$) & 0.7437 & 0.7589 \\
BanditRank-BERT-large ($\gamma =1$) & \textbf{0.7649} &\textbf{0.7807} \\
\hline
\end{tabular}
\end{table}

\subsubsection{Baselines and Results}
We compared the performance of BanditRank against all other previous methods on this dataset. The baselines were CNN-Cnt~\citep{yang2015wikiqa}, QA-CNN~\citep{santos2016attentive}, NASM~\citep{miao2016neural}, Wang et al~\citep{wang2016sentence}, He and Lin~\citep{he2016pairwise}, NCE-CNN~\citep{rao2016noise}, BIMPM~\citep{wang2017bilateral}, Comp-Agg~\citep{wang2016compare}, and the state-of-the-art method Comp-Clip~\citep{bian2017compare}. Since the testing split is same for all methods, we report the highest measures directly from the respective papers.

The results given in Table ~\ref{table: wikiqa results} indicate the superiority of BanditRank over all other methods. BanditRank trained using the features extracted from BERT-large produced the best results. Interestingly in Scenario 1, we observed performance degradation when hybrid loss was used. Although, this degradation may depend on many factors, one possible explanation can be that the model can explore with the help of $L_{rl}$ efficiently since the average number of relevant documents is much less. The results in Scenario 2 indicate that, provided with  good features, training a text-matching network along with the bandit network is not necessary for achieving good results.

\subsection{Experiment 3: MQ2007}
\subsubsection{Dataset and Evaluation Measures}
For the web search task, we used the benchmark Million Query Track 2007 (MQ2007)~\citep{qin2013introducing} dataset~\footnote{https://www.microsoft.com/en-us/research/project/letor-learning-rank-information-retrieval/}. In this task, BanditRank was expected to rank the documents corresponding to a query according to their relevance. Unlike the previous tasks in which the relevance was binary, this task consisted of multiple-levels of relevance with $\{0,1,2\}$. This dataset provides  46-dimensional feature vectors corresponding to each query-document pair. Moreover, the average number of relevant documents per query was very large compared to the previous tasks. The statistics are given in Table~\ref{table: MQ2007}. Out of the total 1692 queries, the number of queries with at least one relevant document was only 1455. 
\begin{table}[h]
\caption{Statistics of MQ2007 dataset}
\label{table: MQ2007}
\begin{tabular}{l c}
\hline
& MQ2007 \\
\hline
\# of queries & 1692 \\
\# of q-with-rel & 1455 \\
\# of documents & 65,323 \\
\# of avg-rel-per-q & 10.3 \\
\# of features & 46\\
\hline
\end{tabular}
\end{table}

The loss for the above mentioned 237 query instances would be zero as the reward generated would be zero. Therefore, we conduct experiments on the dataset after removing the queries with no relevant documents as they would not help BanditRank during training. We carried out 60-20-20 splitting for the train-val-test datasets after the dataset was cleaned. The baselines were also trained and evaluated on the same splits. As per evaluation measures, we report the measures of MAP, MRR, precision, and nDCG at positions 1, 3, 10. We also conducted significance tests using both paired t-test and Wilcoxon signed rank test. 

\subsubsection{Model and Implementation details}
We used Scenario 2 for this task. For the bandit network, we used a feed forward layer with three highway network layers followed by an output layer with a sigmoid unit. The dimensions of the highway layer were set to 92. An input projection layer with an ReLU activation function was used to project the input vectors into 92 dimensions. The provided 46-dimensional feature vectors correspond to vectors $c_i$ mentioned in Section ~\ref{subsec: scenario2}. 

We used the reward function defined in Eq.~\eqref{eq:reward 2}. We chose nDCG@10 instead of reciprocal rank (RR) for this task as the number of relevant documents was large. We optimized the model using the Adam optimizer~\citep{kingma2014adam} with the beta parameters set to $(0,0.999)$, and a weight decay of $1e^{-6}$ was used for regularization. We used the hybrid training objective defined in Eq.~\eqref{eq:hybdrid loss} with $\gamma$ tuned over the set of values $[0.25,0.5,0.75,1]$. The best results were obtained for BanditRank with $\gamma=0.5$. A dropout of 0.4 was applied to all layers. An initial learning rate of $7e^{-5}$ was used for all models. We set $M',B$ to $M'=40$ and $B=30$ for calculating the gradient in Eq.~\eqref{eq:baseline gradient}. A high value was chosen for $M'$ because the number of queries with at least 30 relevant documents was 99, which is a significant number. Moreover, the average number of relevant documents per query was large, i.e., 10.3. 

\begin{table}[h]
\caption{Results of MQ2007 dataset. Best results are in bold. Statistically significant differences compared to best model according to paired t-test is denoted as * and wilcoxon signed rank test is denoted as \textsuperscript{+} (p-value < 0.05).}
\label{table: MQ2007 results}
\begin{tabular}{p{2.65cm}  p{1.1cm} p{1.1cm} p{1.1cm} p{0.7cm}}
\hline
& P@1 & P@3 & P@10 & MAP \\
 \hline
ListNet & 0.446*\textsuperscript{+} & 0.409*\textsuperscript{+} & 0.366*\textsuperscript{+} & 0.452*\textsuperscript{+} \\
AdaRank & 0.474*\textsuperscript{+} & 0.434*\textsuperscript{+} & 0.379*\textsuperscript{+} & 0.471*\textsuperscript{+} \\
Coordinate Ascent & 0.474*\textsuperscript{+} & 0.435*\textsuperscript{+} & 0.382*\textsuperscript{+} & 0.474*\textsuperscript{+} \\
LambdaMART & 0.477*\textsuperscript{+} & 0.444* & 0.390*\textsuperscript{+} & 0.477*\textsuperscript{+} \\
\hline
BanditRank($\gamma=1$) & 0.460	 &0.432 & 0.382 & 0.468 \\
BanditRank($\gamma=0.5$) & \textbf{0.498} & \textbf{0.457} & \textbf{0.393} & \textbf{0.483} \\
\hline
\end{tabular}

\begin{tabular}{p{2.65cm}  p{1.1cm} p{1.1cm} p{1.1cm} p{0.7cm}}
\hline
& nDCG@1 & nDCG@3 & nDCG@10 & MRR \\
\hline
ListNet & 0.391*\textsuperscript{+} & 0.392*\textsuperscript{+} & 0.435 & 0.556*\textsuperscript{+} \\
AdaRank & 0.432*\textsuperscript{+} & 0.426*\textsuperscript{+} & 0.457 & 0.577*\textsuperscript{+}\\
Coordinate Ascent & 0.418*\textsuperscript{+} & 0.420*\textsuperscript{+} & 0.449 & 0.574*\textsuperscript{+} \\
LambdaMART & 0.431*\textsuperscript{+} & 0.434*\textsuperscript{+} & 0.470* & 0.582*\textsuperscript{+}\\
\hline
BanditRank($\gamma=1$) & 0.412	 &0.413 & 0.454 & 0.572 \\
BanditRank($\gamma=0.5$) & \textbf{0.447} & \textbf{0.437} & \textbf{0.473} & \textbf{0.597} \\
\hline
\end{tabular}
\end{table}

\subsubsection{Baselines and Results}
We compared BanditRank with four strong listwise baselines. The baselines were AdaRank~\citep{xu2007adarank}, ListNet~\citep{cao2007learning}, Coordinate Ascent~\citep{metzler2007linear} and the state-of-the-art listwise ranking method LambdaMART~\citep{burges2010ranknet}. All baselines were implemented using the RankLib ~\footnote{https://sourceforge.net/projects/lemur/} software. As mentioned in Section ~\ref{subsec: hybrid training objective}, hybrid training objective with $\gamma=0.5$ resulted in the best performance as BanditRank with  $\gamma=1$ cannot efficiently explore the relatively large action space in this task. 

The results given in Table~\ref{table: MQ2007 results} show that BanditRank clearly outperformed AdaRank, ListNet, and Coordinate Ascent. When compared with the stronger baseline LambdaMART, except for the measures P@10, nDCG@3, and nDCG@10, BanditRank achieved minimum of 1\% improvement in all other measures. Except for the measure nDCG@10, the improvement shown by BanditRank on all other measures is statistically significant according to the paired t-test and wilcoxon signed rank test. In the next section, we discuss the behavior of different reward functions when trained on the MQ2007 dataset. 

\subsection{Comparison of Reward Functions}
\begin{figure}[htb]
\begin{tabular}{cl}
\includegraphics[width=0.24\textwidth]{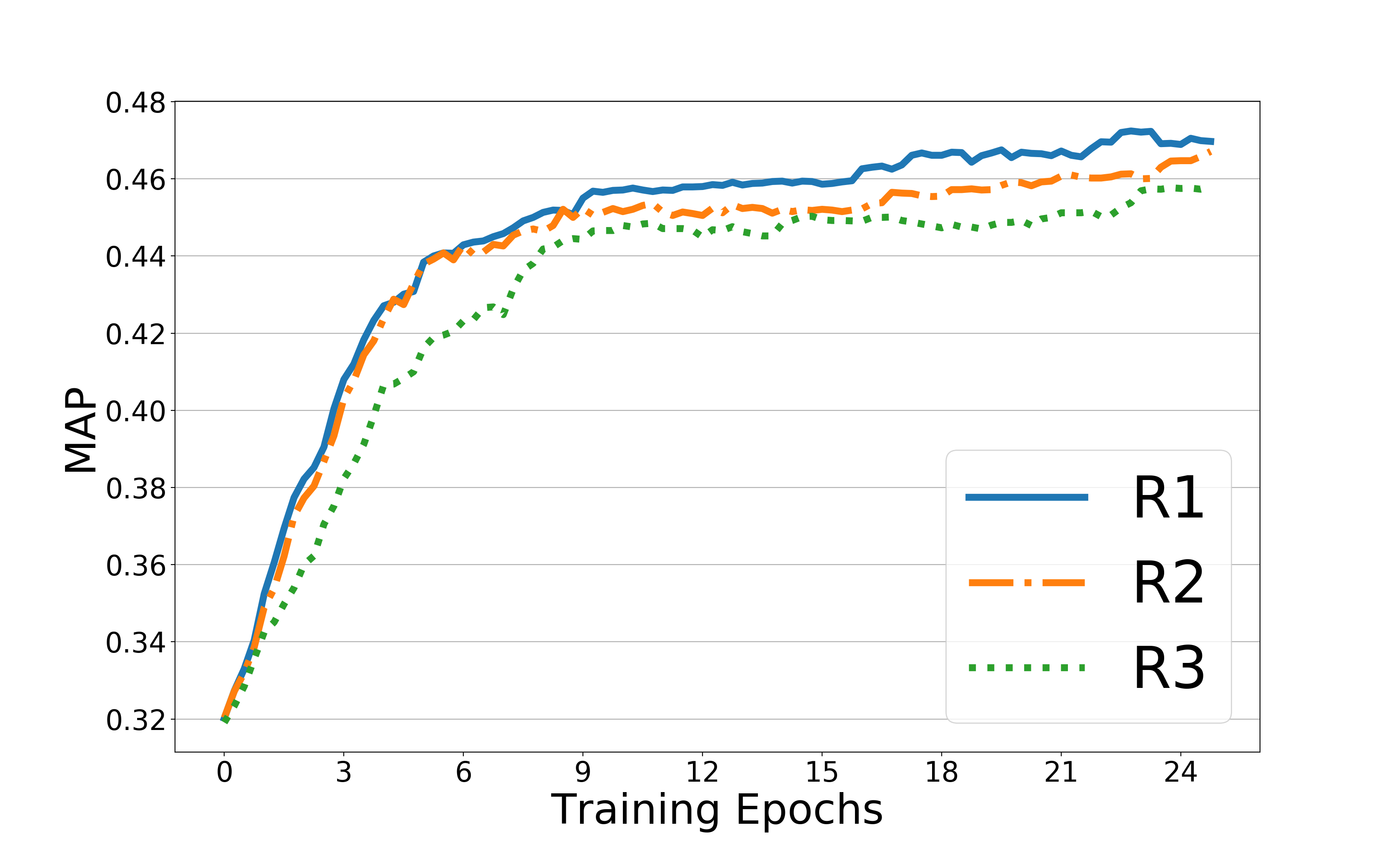} &
\includegraphics[width=0.24\textwidth]{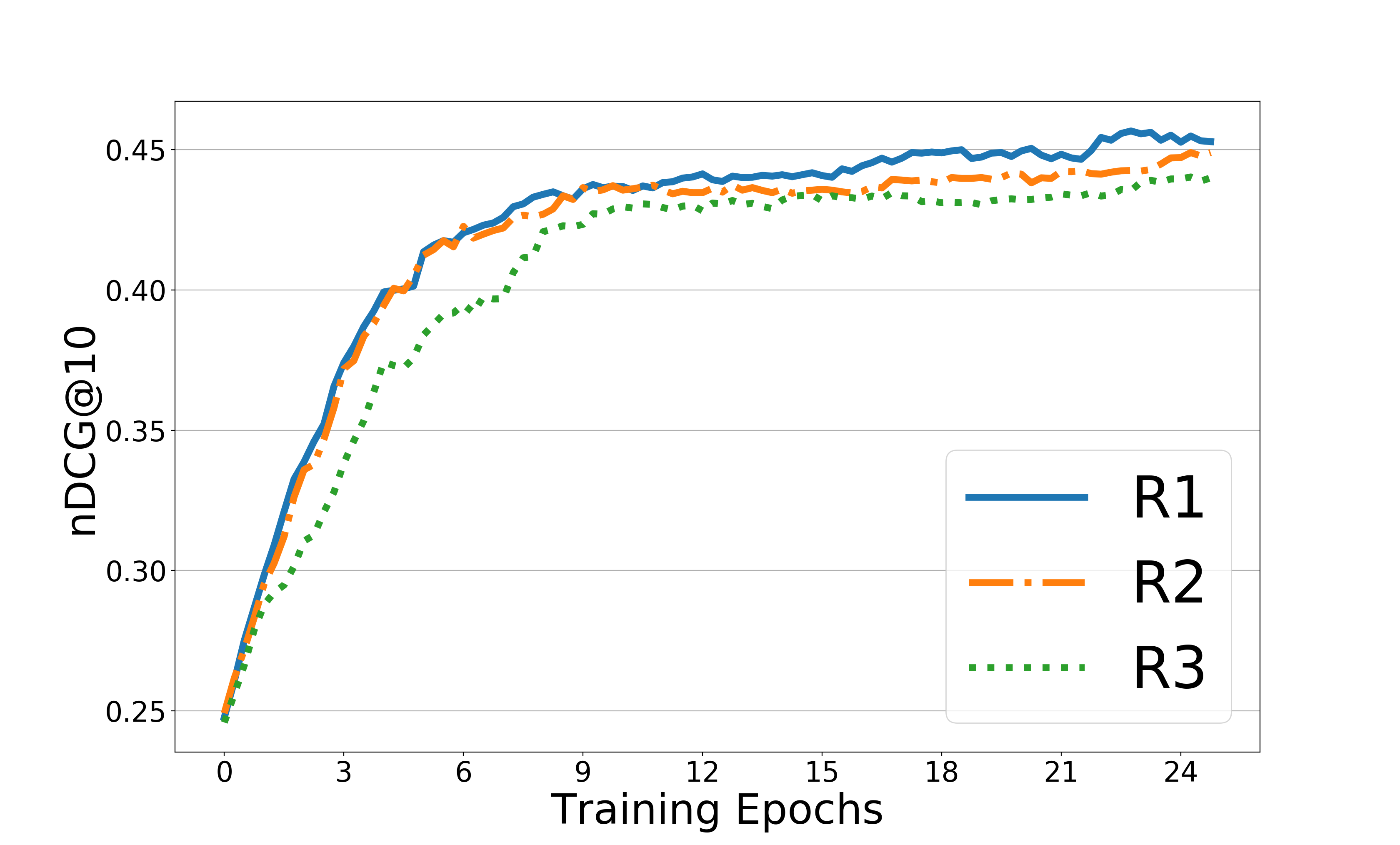} \\
\includegraphics[width=0.24\textwidth]{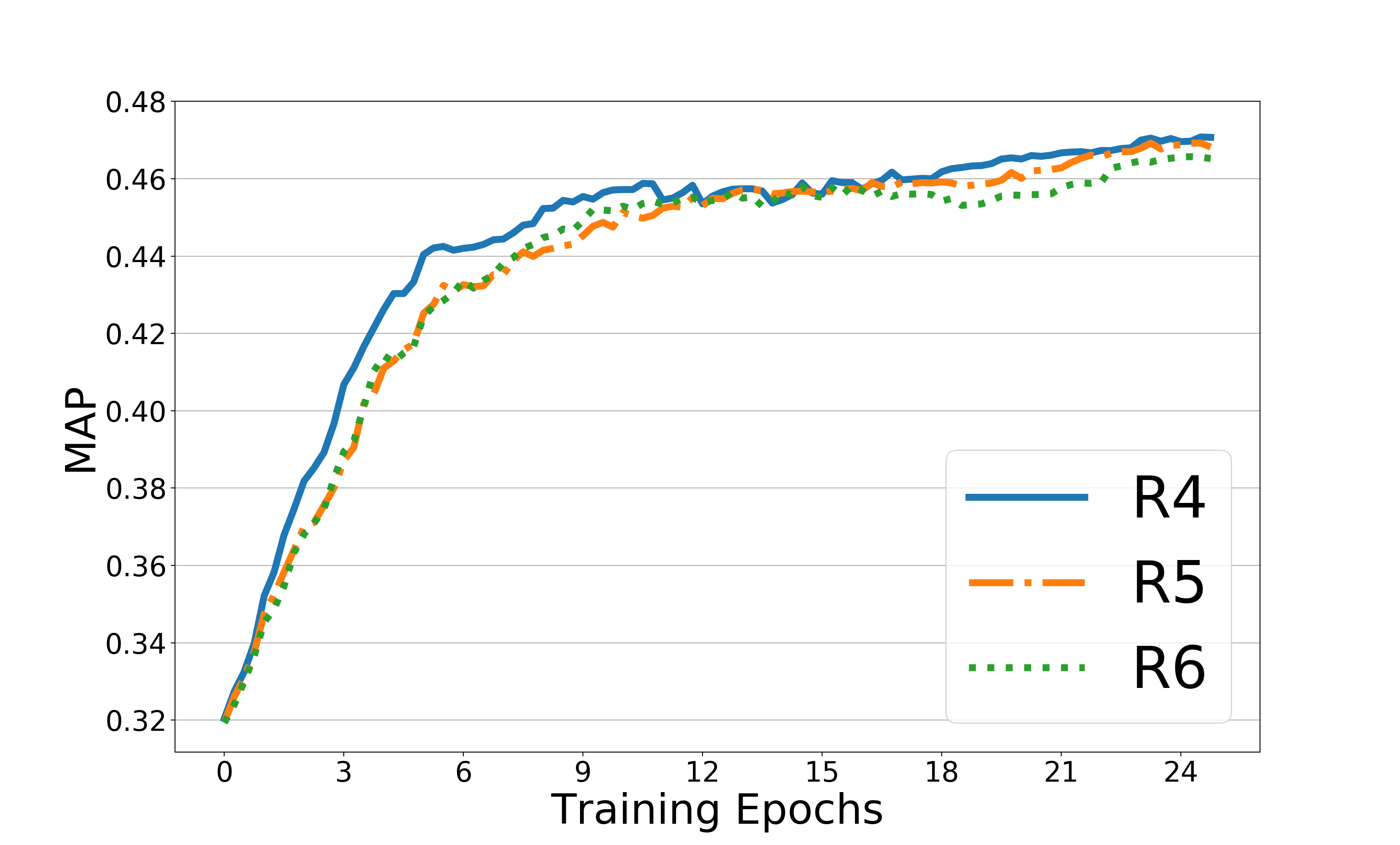} &
\includegraphics[width=0.24\textwidth]{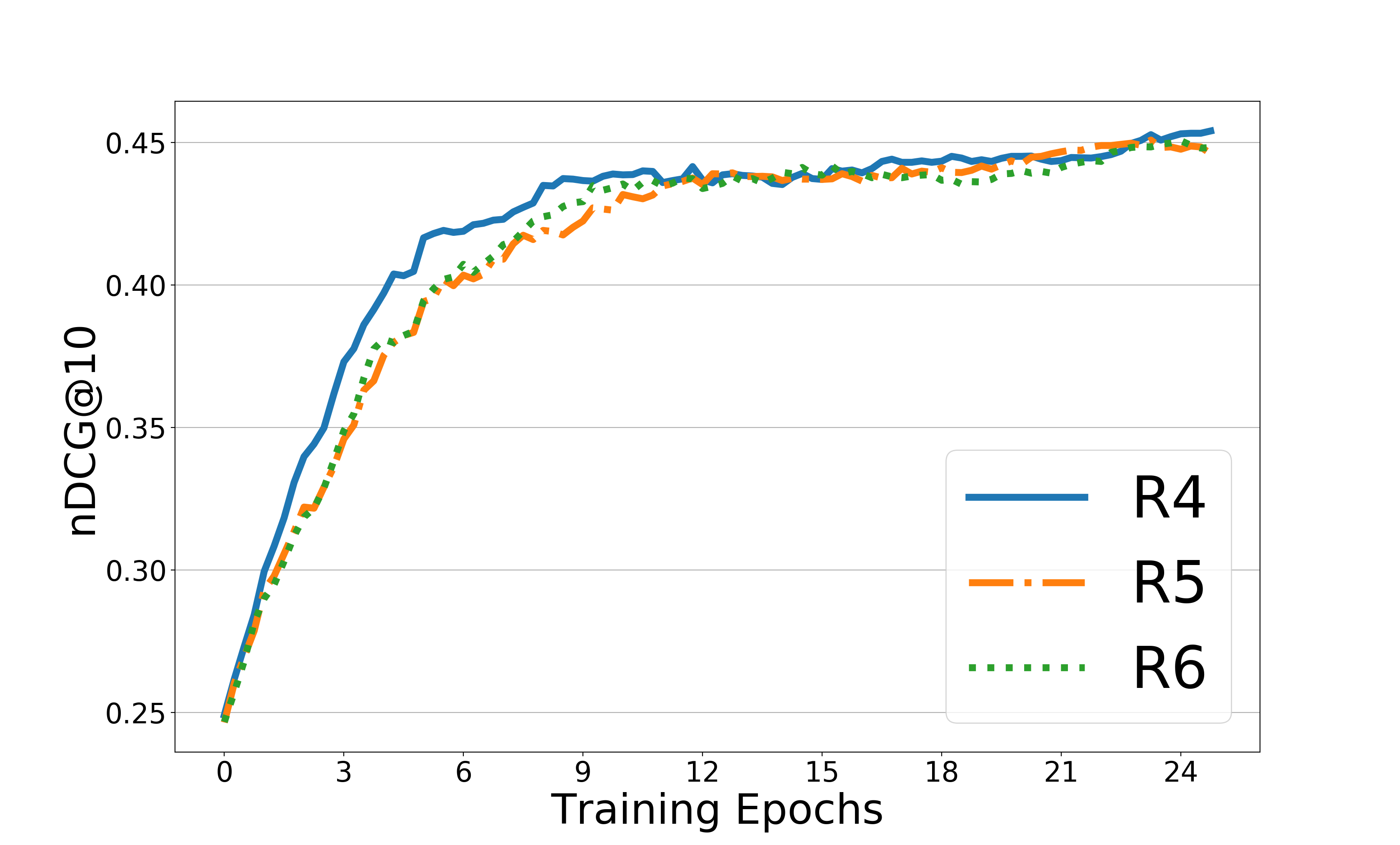} \\
\end{tabular}
\caption{Test set performance of BanditRank trained using different reward functions on MQ2007 dataset. Top row corresponds to functions $R_1, R_2$ and $R_3$ and bottom row corresponds to functions $R_4, R_5$ and $R_6$.}
\label{fig:compare rewards}
\end{figure}

The reward function plays a significant role in the training of an agent in reinforcement learning. Gradual feedback through rewards is often required for training a good agent. For comparing the behavior of reward functions during training, we conducted experiments using different reward functions on the MQ2007 dataset with the same architecture as the previous section. Since we wanted to compare the behavior of the reward function, we chose $\gamma=1$ for all the experiments. The following reward functions were used:
\begin{align*}
R_1(a_c,g_c) &= AP(a_c,g_c)\nonumber \\   
R_2(a_c,g_c) &= nDCG@10(a_c,g_c)  \nonumber \\
R_3(a_c,g_c) &= DCG@5(a_c,g_c) \nonumber \\ R_4(a_c,g_c) &= \frac{[AP+nDCG@10](a_c,g_c)}{2} \nonumber  \\
R_5(a_c,g_c) &= \frac{[AP+RR](a_c,g_c)}{2} \nonumber \\
R_6(a_c,g_c) &= \frac{[AP+P@3+P@5+nDCG@3+nDCG@5](a_c,g_c)}{5}.  \nonumber
\end{align*}
The first three reward functions are the direct evaluation measures and the last three are a combination of several evaluation measures. Figure ~\ref{fig:compare rewards} plots the test set performance of the models during training epochs. We can observe that $R_1$ achieved good measures right from the start when compared with $R_2$ and $R_3$. The performance of $R_6$ when compared with $R_5$ and $R_4$ shows that using many evaluation measures will not necessarily improve the measures of MAP and nDCG@10. During the initial epochs, $R_4$ was clearly a better performer than $R_5$ and $R_6$. After a similar performance by all three models in the next few epochs, $R_4$ achieved better measures of MAP and nDCG@10 in the final stages.

\section{Conclusion}
\label{sec:conclusion}
We proposed an extensible listwise deep learning method \textit{BanditRank} for ranking. It can directly optimize the evaluation measures using the policy gradient algorithm. Experimental results indicate the superiority of BanditRank over other methods on the tested datasets. Future work can involve modifying the structure of the policy network discussed in Section ~\ref{subsec:structure of p} for efficiently addressing the issue of exploration when the number of actions is large. For example, we could use adaptive exploration strategies instead of simple $\epsilon$-greedy strategy, for exploring the action space. We can define new reward functions for handling queries with no relevant documents. For example, we can penalize the model if any of the document affinity scores for such queries is greater than 0.5. There is also a possibility of defining reward functions as the weighted average of different measures with trainable weights for better feedback. Regarding the theoretical aspects, we can compare the directness of BanditRank to other algorithms such as LambdaRank. 
\bibliographystyle{ACM-Reference-Format}
\bibliography{paper}

\end{document}